\documentclass[twocolumn,showpacs,preprintnumbers,amsmath,amssymb,superscriptaddress]{revtex4}

\usepackage{color,latexsym,epsfig,amssymb,amsfonts,amsmath,graphicx,bbm,enumerate,bm}
\usepackage{graphicx}
\usepackage{epstopdf}

\begin{document}

\title{Enhancing multiphoton rates with quantum memories}

\author{J.~Nunn}
\email[]{j.nunn1@physics.ox.ac.uk}
\affiliation{Clarendon Laboratory, University of Oxford, Parks Road, Oxford OX1 3PU, United Kingdom}

\author{N.~K.~Langford}
\affiliation{Department of Physics, Royal Holloway, University of London, Egham Hill, Egham TW20 0EX, United Kingdom}

\author{W.~S.~Kolthammer}
\affiliation{Clarendon Laboratory, University of Oxford, Parks Road, Oxford OX1 3PU, United Kingdom}

\author{T.~F.~M.~Champion}
\affiliation{Clarendon Laboratory, University of Oxford, Parks Road, Oxford OX1 3PU, United Kingdom}

\author{M.~R.~Sprague}
\affiliation{Clarendon Laboratory, University of Oxford, Parks Road, Oxford OX1 3PU, United Kingdom}

\author{P.~S.~Michelberger}
\affiliation{Clarendon Laboratory, University of Oxford, Parks Road, Oxford OX1 3PU, United Kingdom}

\author{X.-M.~Jin}
\affiliation{Clarendon Laboratory, University of Oxford, Parks Road, Oxford OX1 3PU, United Kingdom}
\affiliation{Centre for Quantum Technologies, National University of Singapore, 117543, Singapore}

\author{D.~G.~England}
\affiliation{Clarendon Laboratory, University of Oxford, Parks Road, Oxford OX1 3PU, United Kingdom}

\author{I.~A.~Walmsley}
\affiliation{Clarendon Laboratory, University of Oxford, Parks Road, Oxford OX1 3PU, United Kingdom}

\date{\today}
\begin{abstract}
Single photons are a vital resource for optical quantum information processing. Efficient and deterministic single photon sources do not yet exist, however. To date, experimental demonstrations of quantum processing primitives have been implemented using non-deterministic sources combined with heralding and/or postselection. Unfortunately, even for eight photons, the data rates are already so low as to make most experiments impracticable. It is well known that quantum memories, capable of storing photons until they are needed, are a potential solution to this `scaling catastrophe'. Here, we analyze in detail the benefits of quantum memories for producing multiphoton states, showing how the production rates can be enhanced by many orders of magnitude. We identify the quantity $\eta B$ as the most important figure of merit in this connection, where $\eta$ and $B$ are the efficiency and time-bandwidth product of the memories, respectively.
\end{abstract}


\maketitle
After two decades of rapid advances, quantum optics experiments are becoming increasingly challenging. As the interests of the community shift to higher-dimensional entanglement \cite{Lanyon:2009uq,Wieczorek:2009kx} and information processing tasks beyond mere proof-of-principle \cite{Thompson:2011kx,Thomas-Peter:2011vn}, the demand for large numbers of simultaneous single photons is outstripping the capabilities of parametric sources \cite{Cohen:2009hc,Mosley:2008hs}. These sources, which so far have been the workhorse of the quantum optics lab, produce photons in pairs, but they also produce multiple unwanted photon-pairs with a probability that scales with the single-pair generation rate, which must therefore be kept low, so that most often no photons are emitted. The current record for photonic resources is an eight-photon experiment involving four parametric sources, in which statistics were accumulated over 40 hours \cite{Yao:2011fk}. The approach of accessing higher photon numbers by simply waiting longer is not sustainable.

In this paper we study the use of quantum memories to store and synchronise randomly-emitted heralded single photons as a means to efficiently construct the multiphoton states needed for quantum information processing with light \cite{Hosseini:2011zr,Reim:2011ys}. The development of quantum memories has historically been motivated by applications in quantum communications, where signals traverse large distances and long storage lifetimes are therefore required \cite{Sangouard:2008xr}. While no currently available memory has the performance characteristics necessary for these demanding communications applications, we find that even inefficient memories, which are available now, can enhance the generation rate of multiphoton states by enormous factors. In this context, long storage times are not necessary, but a large time-bandwidth product is a key parameter.


 The photonics scalability problem is easily understood. Suppose that a single photon is heralded with probability $q\ll1$. The probability of producing $N$ single photons simultaneously using $N$ sources is then simply $q^N$, which becomes exponentially small as $N$ increases, thus rendering complex experiments impossible.
 
 Some kind of multiplexing strategy is necessary to mitigate this problem. In \emph{spatial multiplexing}, sources are operated in parallel and their outputs are combined via active switching \cite{Migdall:2002cr,OBrien:2009dq,Jennewein:2011zr,Ma:2011ly,McCusker:2009fk,Hall:2011qf}. Many identical sources are required to achieve efficient operation \cite{Christ:2012bh}. This may be achievable with emerging integrated optics platforms \cite{Matsuda:2012uq,Fortsch:2012kx}, but it is well-known that \emph{temporal multiplexing}, where sources are operated many times in series and their outputs are synchronised using quantum memories, offers an alternative solution \cite{Yuan:2007vn, Pittman:2002ys, Mower:2011nx}. In general, a hybrid approach involving both spatial and temporal multiplexing could be adopted. Here we consider how temporal multiplexing allows to re-time probabilistic sources and boost the $N$-photon generation rate.
 
\begin{figure}[h]
\begin{center}
\includegraphics[width=5cm]{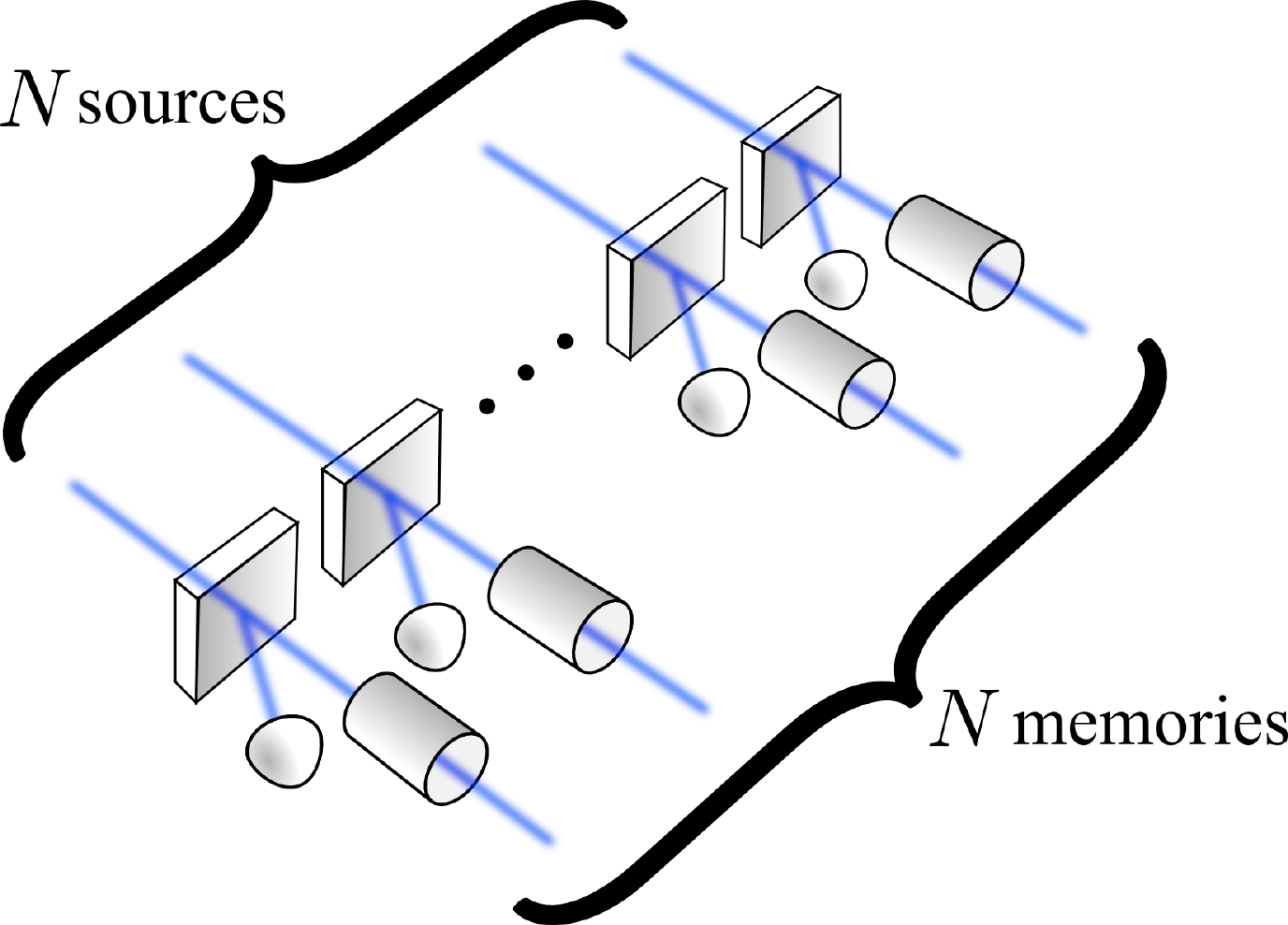}
\caption{An array of $N$ heralded parametric sources synchronized by quantum memories. The sources are repeatedly pumped, and each photon emitted is stored until all but one of the memories are charged. Then emission of a photon by the final source triggers retrieval from the memories, in order to generate an $N$-fold coincidence.}
\label{fig:array}
\end{center}
\end{figure}

 To see how quantum memories can increase the rate of $N$-fold coincidences, consider the array of $N$ sources coupled to $N$ memories shown in Fig.~\ref{fig:array}. We suppose that each source produces photons in pairs by means of a parametric scattering process such as downconversion \cite{Mosley:2008hs} or spontaneous four-wave mixing \cite{Cohen:2009hc}, with one of each pair directed to a herald detector. With no memories, all $N$ heralds must fire simultaneously to produce an $N$-fold coincidence. However with memories, heralded photons can be stored whenever they are produced. Once $N-1$ of the memories are charged with a photon, one only has to wait for the final source to produce a photon, and then all the memories can be read out and one has, again, an $N$-fold coincidence. This protocol may not optimal, but it is amenable to a straightforward analysis that captures the scaling enhancement: by lifting the requirement for simultaneous emission, the memories greatly enhance the coincidence probability. Our purpose in this paper is to quantify the gain in coincidence rate afforded by using quantum memories to synchronize photon sources in this way. The time-bandwidth product $B = \delta \tau$ proves to be critical in this context, where $\delta$ is the acceptance bandwidth of the memories, and $\tau$ is their coherence lifetime \footnote{This is not the same as the multimode capacity \cite{Nunn:2008oq,Simon:2007ct}.}. If postselection on the final detection of $N$ photons is used, even relatively inefficient memories can dramatically enhance the multiphoton rate.

Some additional assumptions are needed to fully specify the protocol. First, we assume that we always attempt to store a photon, if a herald detector fires, regardless of whether or not the memory concerned is already charged. This ensures that we always use the most recently emitted photons, which mitigates photon loss due to decoherence in the memories. Second, to avoid `clashes' (in fact, interference \cite{Campbell:2011kl,Datta:2012tg}) between incident and stored photons, we also assume that we clean each memory before storage is attempted (\emph{e.g.} by readout of the memory, or optical pumping) so that we are always attempting to charge an empty memory. This allows us to use a classical model, in which individual photons are treated as particles that are probabilistically emitted, stored and retrieved. Finally, we adopt the policy that if we are ready to read out the memories --- that is, if $N-1$ memories have been charged and a photon is emitted from the $N^\mathrm{th}$ source --- and at the same time one or more of the other sources emits a photon, we bypass the relevant memories and use these `serendipitous' photons, rather than attempting to read out the memories.

The photon sources are pumped at a rate $\mathcal{R}\sim \delta$, limited by the minimum pulse duration that can be stored by the memories. The average waiting time $1/\mathcal{R}c_\mathrm{sync}$ between $N$-photon events can then be computed if we can find an expression for the $N$-fold coincidence probability $c_\mathrm{sync} = p_\mathrm{sync}^N$, which is the probability that one photon is obtained from each of the $N$ source-memory units. We have defined $p_\mathrm{sync}=q+\overline{q}\eta_\mathrm{r}P$ as the probability that any source-memory unit provides a photon on demand, either directly, or through successful retrieval of a stored photon. Here $P$ is the steady-state probability that any memory is charged with a stored photon, $\eta_\mathrm{r}$ is the retrieval efficiency, and the overbar notation denotes the probabilistic complement, $\overline{X}\equiv1-X$. The problem of computing the waiting time then reduces to that of finding $P$. To proceed we assume that the decoherence processes in the memories are Markovian (\emph{i.e.} exponential), since then the stochastic evolution of the charge-state $\bm{x}^{(m)} = [\overline{P}^{(m)},P^{(m)}]^\mathsf{T}$ of each memory can be tracked using a transfer matrix:
\begin{equation}
\label{T}
\bm{x}^{(m)} = T \bm{x}^{(m-1)};\qquad T= \left(\begin{array}{cc} \overline{r} & s \\ r & \overline{s}\end{array}\right),
\end{equation}
with $P^{(m)}$ the probability that the memory is charged at the $m^\mathrm{th}$ time step, $r$ the probability that an empty memory becomes charged over the course of one time step, and $s$ the probability that a charged memory becomes empty. The steady state probabilities are given by the eigenvector $\bm{x}_\mathrm{s}$ of $T$ with eigenvalue $1$, $\bm{x}_\mathrm{s} = [s, r]^\mathsf{T}/(r+s)$, so that we have $P = r/(r+s)$. The probability that an empty memory becomes charged is the probability that a heralded photon is emitted and that it is stored, provided that the rest of the set-up is not primed for readout, so we have $r = q\eta_\mathrm{s}\overline{R}$, where $R$ is the probability that the system is ready to be read out (the evaluation of $R$ is described in the Appendix), and $\eta_\mathrm{s}$ is the storage efficiency. Note that this storage efficiency can include any coupling or propagation losses between the source and the memory input. The loss probability that a charged memory is emptied is more complicated. The probability of decoherence during any time step is $b=1-e^{-1/B}$, where generally the time-bandwidth product will be much larger than one, so that $b \approx 1/B$. There are then four loss processes to consider. First, decoherence in the memory during standby ($b\overline{q}\overline{R}$), second, decoherence in the memory during the readout stage, when a photon is heralded and the memory is bypassed, leaving the memory charged and vulnerable to decay ($bqR$), third, readout of the memory when we attempt to generate a coincidence ($R\overline{q}$), and finally, the loss of a stored photon during standby when a new photon comes along and we attempt to replace the stored photon but fail ($q\overline{\eta}_\mathrm{s}\overline{R}$). The total loss probability is then $s=b[\overline{R}\overline{q} +Rq] + \overline{q}R + q\overline{R}\overline{\eta}_\mathrm{s}$, and we finally obtain
\begin{equation}
\label{multiphoton}
c_\mathrm{sync}=q^N\left\{1+\frac{\overline{R}\overline{q}\eta B}{1+(B-1)\left[R(\overline{q}-q)+q\right]}\right\}^N.
\end{equation}
In the limit of small photon generation rates such that $\{RB,qB\}\ll1$, we have $c_\mathrm{sync}\approx (q\eta B)^N$, which supports the intuition that each memory effectively boosts the photon generation probability by $B$, moderated by its efficiency $\eta=\eta_\mathrm{s}\eta_\mathrm{r}$. In this regime the gain in the multiphoton rate is therefore exponential in $N$ with the base quantity $\eta B$, which highlights the importance of the time-bandwidth product for synchronization applications. As $B$ is increased so that $qB\gg1$, the rate eventually saturates and becomes independent of $B$, limited finally by $\eta$. Note that if the heralding detectors suffer dark counts with probability $d\ll1$, one simply replaces the leading factor of $q^N$ in Eq.~(\ref{multiphoton}) with $(q-d)^N$.

 To make a fair comparison with the unsynchronized case, we now consider the effect of higher photon number components on the quality of the states produced. Typically, parametric sources generate photon pairs according to a thermal distribution, where $p_\mathrm{source}(n) = \overline{p}p^n$ is the probability of emitting $n$ photon pairs, and $p$ is a small real number. We also assume non-photon-number-resolving heralding detectors, such as APDs, so that the conditional probability that $n$ photons are sent towards a memory, given a herald click, is $p_\mathrm{h}(n)=\overline{d}[1-\overline{h}^n+d]p_\mathrm{source}(n)/q$ where $h$ is the efficiency of the heralding detector, and as before $q = (hp+d\overline{p})/[1-p\overline{h}]$ is the probability of a herald click. The charge state $\bm{x}^{(m)}$ of the memory is now a vector of probabilities that the memory contains $n$ photons, with $n=0,1,2...$, which we truncate for numerical convenience. The transition probability that the number of excitations stored in a memory changes from $k$ to $j$ over the course of any time step is given by the transfer matrix element
\begin{equation}
\label{T2}
T_{jk}= \theta_{jk}b^{k-j}\overline{b}^j{{k}\choose{j}}(\overline{R}\overline{q}+Rq)+ \overline{q}R\delta_{j0}+q\overline{R}p_\mathrm{s}(j),
\end{equation}
where the three terms represent decoherence, readout, and storage, respectively. Here $\theta_{jk}=1$ for $k\geq j$ and zero otherwise (since excitations are only lost through decoherence), and $\delta_{j0}$ is a Kronecker delta that describes the erasure of the memory after a read-out event. We have also defined $p_\mathrm{s}(n)$ as the probability that $n$ photons are stored in the memory when read-in is attempted after a herald, $p_\mathrm{s}(n)=\sum_{k=n}^\infty p_\mathrm{h}(k)\eta_\mathrm{s}^n\overline{\eta}_\mathrm{s}^{k-n}{{k}\choose{n}}$. Repeated application of $T$ to an arbitrary initial charge state converges to the steady state $\bm{x}_\mathrm{s}$, and the probability that $n$ photons are retrieved from the memory is then given by $p_\mathrm{r}(n) = \sum_{k=n}^\infty x_\mathrm{s}(k)\eta_\mathrm{r}^n\overline{\eta}_\mathrm{r}^{k-n}{{k}\choose{n}}$. We can then write the $N$-fold coincidence probability as $c = p_\mathrm{sync}(1)^N$, where $p_\mathrm{sync}(n)=qp_\mathrm{h}(n) + \overline{q}p_\mathrm{r}(n)$ is the probability that $n$ photons are successfully extracted from each of the source-memory units. This result for $c$ represents only a minor correction to Eq.~(\ref{multiphoton}), but the treatment of multi-pair emissions is important for the fidelity calculation below.

To compare synchronised and unsynchronised systems, we calculate the fidelity of the $N$-mode states they produce with the ideal state of exactly one photon in each mode. The fidelity decreases exponentially with $N$, so that according to this measure, even very high quality sources would have low fidelity for large $N$. To capture the performance of individual sources synchronised with our scheme, we therefore normalise the fidelity to the number of modes by taking the $N^\mathrm{th}$ root. This yields a measure $\mathcal{F}$ that does not decay exponentially with $N$, which can then be applied to benchmark individual components of a large device against some error correction threshold.

For the synchronised system, we obtain $\mathcal{F}_\mathrm{sync} = \left[c_\mathrm{sync}/RY\right]^{1/N}$, where $RY$ is the probability that we believe we have produced an $N$ photon state (see Appendix). Without memories, the $N$-fold coincidence rate is $c_\mathrm{no\;mem}=[qp_\mathrm{h}(1)]^N$ and the normalised fidelity is simply given by $\mathcal{F}_\mathrm{no\;mem}=p_\mathrm{h}(1)$. Note that here we have ignored any losses (such as fibre-coupling losses) that could reduce $\mathcal{F}_\mathrm{no\;mem}$, and so the results below are a conservative estimate of the benefits of synchronisation.

In many photonic networks, successful operations can be postselected on the final detection of at least $N$ photons. In this case the fidelity of the postselected states is the fraction of these which contain $1$ photon per mode, and the normalised postselected fidelity $\widetilde{\mathcal{F}}$ of the unsynchronised system is simply $\widetilde{\mathcal{F}}_\mathrm{no\;mem}= \mathcal{F}_\mathrm{no\;mem}$, since we assumed that the heralding already completely removed the vacuum component. For the synchronised system, the normalised postselected fidelity is
\begin{equation}
\label{post_fid}
\widetilde{\mathcal{F}}_\mathrm{sync} = \left[\frac{c_\mathrm{sync}}{p_{\geq N}}\right]^{1/N}=\left[\frac{c_\mathrm{sync}}{q-p_{<N}}\right]^{1/N}.
\end{equation}
Here $p_{\geq N}$ is the probability that the state obtained from the memories/sources comprises $N$ photons or more, and we have re-written this in terms of $p_{<N}$, the probability that fewer than $N$ photons in total are emitted, given by
\begin{equation}
\label{p_less}
p_{<N}=\sum_{j=0}^{N-1}\sum_{\bm{s}_j} \prod_{l=1}^{N} p_\mathrm{sync}(s_j(l)).
\end{equation}
Here $s_j(l)$ is the $l^\mathrm{th}$ element of a vector $\bm{s}_j$ containing $N$ real, non-negative integers whose sum is equal to $j$. The summation $\sum_{\bm{s}_j}$ runs over all such vectors.

\begin{figure}[h]
\begin{center}
\includegraphics[width = \columnwidth]{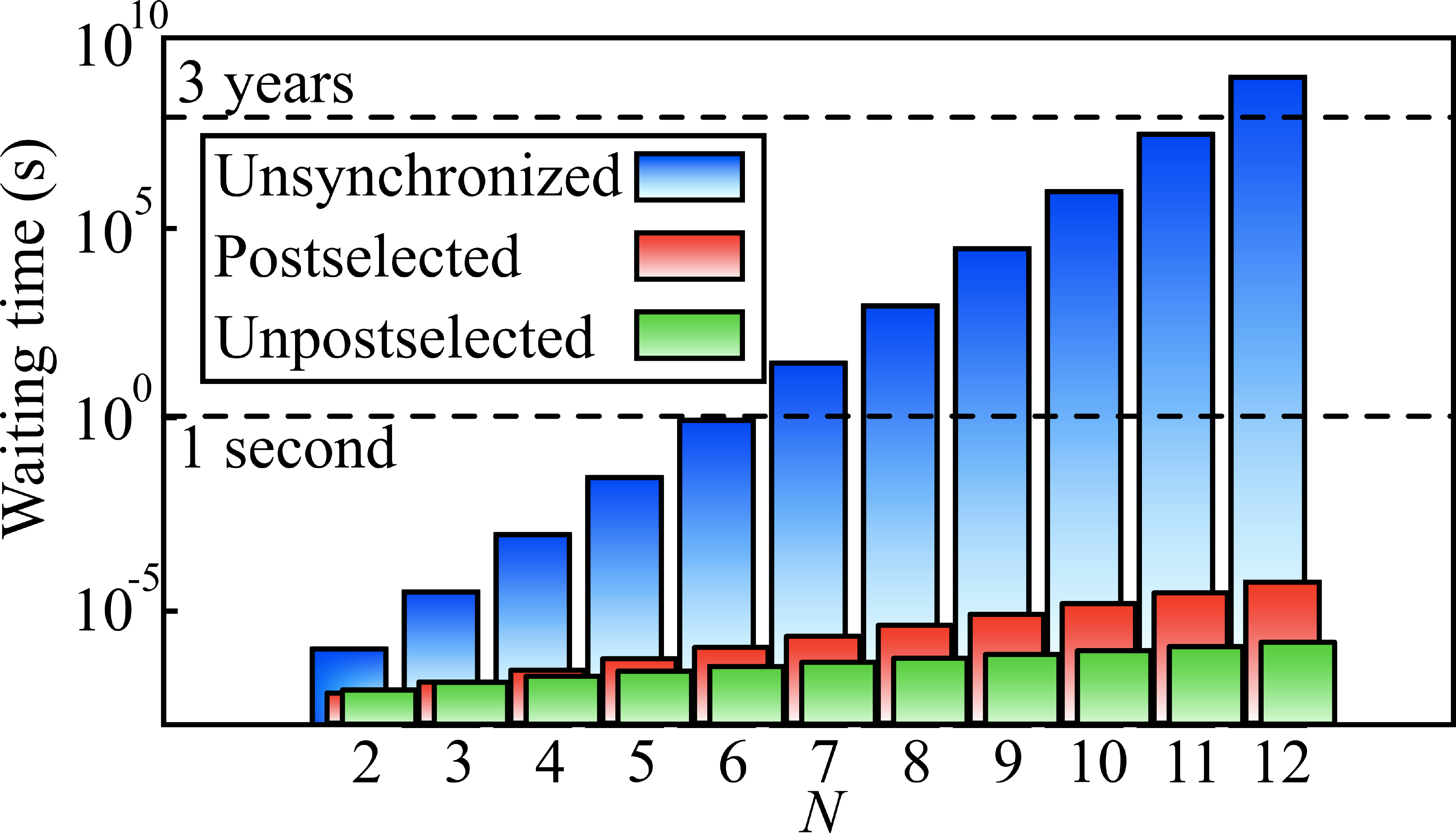}
\caption{Multiphoton waiting times. The blue bars show the average waiting time between $N$-photon events for a system of $N$ unsynchronized downconversion sources, assuming negligible dark counts ($d=0$), a pulse repetition rate of $\mathcal{R}=1$~GHz, a heralding efficiency of $h=50\%$ and a threshold fidelity $\Theta = 90\%$. The red bars show the corresponding waiting times when the system is synchronized, with memory efficiencies $\eta_\mathrm{s}=\eta_\mathrm{r}=75\%$ and a time-bandwidth product $B=1000$, where postselection on at least $N$ photons is used (we set $p_\Theta$ so that $\widetilde{\mathcal{F}}=\Theta$). The green bars show the waiting times without postselection (we set $p_\Theta$ so that $\mathcal{F}=\Theta$), where to achieve the required unpostselected fidelity threshold we now assume memory efficiencies of $\eta_\mathrm{s}=\eta_\mathrm{r}=99\%$.}
\label{fig:waiting_times}
\end{center}
\end{figure}

In general, neither postselected nor unpostselected fidelities for either synchronized or unsynchronized systems will reach 1, except in the limit $p\rightarrow0$. Therefore one must choose a threshold fidelity $\Theta$ that is acceptable, and then one should choose the largest value $p_\Theta$ of $p$ such that $\mathcal{F},\widetilde{\mathcal{F}}=\Theta$, for each system. Having done this, one can then compare the $N$-fold coincidence rates. Assuming for simplicity $d=0$, we have $p_\Theta = \{2-h-[(2-h)^2-4\overline{h}\overline{\Theta}]^{1/2}\}/2\overline{h}$ for an unsynchronised system, independent of $N$. For the same number of synchronized sources, $p_\Theta$ depends on $N$ and needs to be determined by a numerical optimisation. Figure~\ref{fig:waiting_times} shows the resulting comparison of synchronized and unsynchronized systems.
The waiting times scale exponentially with the number $N$ of photons required, and without synchronization a 12 photon experiment would require more than 30 years in between coincidence events, so that quantum computing with photons using such a system is totally unfeasible. However the use of memories reduces the waiting time quite dramatically. For a postselected experiment one can use inefficient memories with $\eta = 56\%$ ($\eta_\mathrm{s}=\eta_\mathrm{r}=75\%$) and reduce the 12-fold waiting time to $\sim$100~$\mu$s. Quantum memories based on Raman scattering have already been demonstrated with $\delta>$~GHz \cite{Reim:2010kx}, $B>1000$ \cite{Reim:2011ys,Bao:2012fk} and $\eta > 50\%$ \cite{Hosseini:2011zr}, while highly efficient and multiplexed storage in rare-earth memories is maturing \cite{Hedges:2010kx,Usmani:2010uq, Afzelius:2010fk, Saglamyurek:2011fk, Simon:2007ct}, and so these dramatic enhancements lie well within the reach of current technology. Without postselection more efficient memories are required to achieve the fidelity threshold ($\eta_\mathrm{s}=\eta_\mathrm{r}=99\%$), which puts the implementation beyond current technological capabilities since $\eta=98\%$ efficiency has not yet been demonstrated. But if this improved performance were achieved, the waiting time would be further reduced to $\lesssim$1~$\mu$s.

In summary, we have analyzed the use of quantum memories for the synchronization of multiple single photon sources as a canonical application of quantum storage for the enhancement of photonic information processing. We derived an analytic formula for the multiphoton rate achievable and showed that the most important figure of merit for quantum memories is the product $\eta B$ of the memory efficiency with its time-bandwidth product. Finally we extended our model to include higher-order photon number contributions, so that the quality of the states produced with and without memories could be compared. We showed that even inefficient memories can produce enormous improvements in the multiphoton rate when combined with postselection. Without postselection, highly efficient memories are required to match the quality of unsynchronized sources, but if these are available the gain in multiphoton rate becomes larger still. It would be interesting to consider the effects of noise in the memories, or extensions to more complicated synchronization protocols. It is expected that similar advantages could pertain to the scaling of other heralded quantum operations, such as entanglement generation or two-photon gates. While much attention in the quantum memory community has focussed on the need for long storage times and high efficiencies in the context of quantum repeaters \cite{Duan:2001vn,Sangouard:2008xr}, our analysis underlines the value of developing quantum memories for local synchronization, for which lower efficiencies still provide considerable advantages, and for which the time-bandwidth product $B$ is much more important than the absolute storage time.
\vspace{-.5cm}
\acknowledgements
\vspace{-.5cm}
This work was supported by the EPSRC (EP/C51933/01), the EC project Q-ESSENCE (248095), the Royal Society, and the AFOSR EOARD. XMJ acknowledges support from NSFC (11004183) and CPSF (201003327). NKL was supported by an EC Marie-Curie Fellowship. MRS was supported by a Clarendon Scholarship. PSM was supported by the EC ITN FASTQUAST. JN is grateful to Marco Barbieri for wearing him down.
\vspace{-.5cm}
\appendix
\vspace{-.5cm}
\section*{Appendix}
\vspace{-.5cm}
The readout probability $R$ can be computed by tracking the belief state $\bm{y}^{(m)} = [\overline{V}^{(m)},V^{(m)}]^\mathsf{T}$, where $V^{(m)}$ is the probability that we believe the memory to be charged at the $m^\mathrm{th}$ time step. We have
\vspace{-.2cm}
$$
\bm{y}^{(m)} = S \bm{y}^{(m-1)};\qquad S = \left(\begin{array}{cc} \overline{w} & z \\ w & \overline{z}\end{array}\right),
\vspace{-.2cm}
$$
where $w=\overline{R}q$ ($z = \overline{q}R$) is the probability that we believe an empty (charged) memory becomes charged (empty) over the course of one time step. In the steady state $V^{(m)}\rightarrow V = w/(w+z)$. On the other hand, readout occurs when we believe $N-1$ other photons to be available, so we can write $R = Y^{N-1}$, where $Y=q+\overline{q}V$ is the probability that we believe a source has provided a photon, either directly or through its memory. Combining these relations we obtain the consistency condition $(1-2q)Y^N + q^2 Y^{N-1} + Yq - q = 0$, the positive real root of which can be found numerically, which then fixes $R$.

\end{document}